\documentclass[aps,pre,floats,twocolumn,showpacs,superscriptaddress]{revtex4}

\usepackage{epsfig}
\usepackage{latexsym,amsmath}

\begin{document}
\title{Generalized percolation in random directed networks}

\author{Mari{\'a}n Bogu{\~n}{\'a} and M. \'Angeles Serrano}

\affiliation{Departament de F{\'\i}sica Fonamental, Universitat de
  Barcelona, Mart\'{\i} i Franqu\`es 1, 08028 Barcelona, Spain}

\date{\today}

\begin{abstract}
We develop a general theory for percolation in directed random networks
with arbitrary two point correlations and bidirectional edges, that is, edges pointing
in both directions simultaneously. These two ingredients alter the previously known
scenario and open new views and perspectives on percolation phenomena. Equations for the percolation threshold and the sizes of the giant components are derived in the most general case. We also present simulation results for a particular example of uncorrelated network with bidirectional edges confirming the theoretical predictions. 
\end{abstract}


\maketitle

\section{Introduction}

A wide class of real systems of many interacting elements can be
mapped into graphs or networks. Under this approach, vertices or
nodes of the network represent the elements of the system whereas
edges or links among them stand for interactions between different
elements. This mapping has triggered a huge number of works and a
surge of interest in the field of complex networks that has lead
to a general framework within which to analyze their topology as
well as the dynamical processes running on top of them \cite{Mendesbook,Dorogovtsev02,Barabasi02}. In many
cases, these dynamical processes are directly related to
functionality and involve some kind of transport or traffic flow.
Furthermore, the very existence of those networks could be
naturally explained as a direct consequence of the communication
need among its constituents. The Internet or the World Wide Web
are clear examples \cite{Romusbook}. In order to preserve functionality, networks
characterized by transport processes must be connected, that is, a
path must exist between any pair of nodes, or, at least, there
must exist a macroscopic portion of vertices --or giant
component-- able to communicate. In this context, percolation
theory appears as an indispensable tool to analyze the conditions
under which such connected structures emerge in large networks.

The general theory of percolation phenomena for uncorrelated
undirected random networks was first developed by Newman {\it et
al.} \cite{Newman01,Callaway00} after previous results in
\cite{Molloy95,Molloy98}. The phase transition at which the giant
component forms was well characterized, and the size distribution of connected finite components below and
above the critical point were calculated as well. Some further
refinements were needed in order to approach real
nets. Hence, correlations between degrees of neighboring vertices
were taken into account in \cite{Newman022,Newman03,Vazquez03} and growing networks
were treated by Dorogovtsev {\it et al.} \cite{Dorogovtsev01b} and Krapivsky {\it et al.} \cite{Krapivsky04}.

To go further, directness must be taken into consideration since
some of the most interesting real networks present asymmetric
interactions. Noticeable examples are the World Wide Web
\cite{Broder00}, citation networks \cite{Price65,Redner98}, email networks
\cite{Newman02}, gene regulatory networks \cite{Jong02}, or metabolic
networks \cite{Fell00}. Percolation theory for purely directed networks was
first developed by Newman {\it et al.} \cite{Newman01,Callaway00}
and later by Dorogovtsev {\it et al.} \cite{Dorogovtsev01}. In the particular case of scale-free
degree distributions a number of interesting specific results were
obtained \cite{Schwartz02}. While allowing for general
correlations between the incoming and the outgoing number of edges
of a given vertex, all these studies refer to networks with no
degree correlations. Furthermore, the theory is restricted to the
class of directed networks with no bidirectional edges, although this class of edges are ubiquitous to all real directed networks (see \cite{Garlaschelli04} and references therein). All these
limit the applicability of the theory to real networks since
bidirectional edges and degree correlations are common to all real
directed networks. In this paper we present a general theory for
percolation in directed random networks with general two point
correlations and bidirectional links. We will show that the
presence of bidirectional edges and degree correlations modify the
picture previously drawn
\cite{Newman01,Callaway00,Dorogovtsev01,Schwartz02} in a non
trivial way, opening new scenarios for percolation phenomena.

The paper is organized as follows. In section II, we review
concepts, definitions, and the main results previously obtained in
the analysis of percolation in random networks. In section III, we
develop the general theory for directed networks with
bidirectional edges and arbitrary degree correlations and we show
how the theories for undirected and purely directed random
networks stand as particular cases. Section IV is devoted to the
uncorrelated case, which deserves special attention as a null or
benchmark model. The relative sizes of the giant components are
computed and the explicit expression for the percolation condition
is provided. The well-known critical points signalling the phase
transition in undirected and purely directed uncorrelated networks
are recovered as limiting cases. A practical application of the
formalism for uncorrelated networks is presented in section V,
where the transformation from a purely directed network to a
purely bidirectional one is studied. For power-law degree
distributions, the transformation is shown to undergo a nontrivial
phase transition. Simulation results support this prediction, finding an 
excellent agreement with numerical solutions of the theoretical
equations. Finally, we conclude with a brief report of results in
section VI.

\section{Percolation in uncorrelated purely directed networks}

The topological structure of directed networks is more complex
than that associated to undirected graphs. The edges associated to
each node in a directed net are usually differentiated into
incoming and outgoing. Usually, no bidirectional links are
considered so that each vertex has two coexisting degrees, $k_i$
and $k_o$, which sum up to the total degree $k=k_{in}+k_{out}$.
Hence, the degree distribution for a directed network is a joint
degree distribution $P(k_{i},k_{o})$ of in- and out-degrees, which
in general may be correlated. The bidirectional edge symmetry of
undirected networks is thus completely broken in purely directed ones,
with implications down to the level of percolation properties. The
giant connected component in undirected graphs becomes internally
structured in the case of directed networks so that four different
types of giant components may arise. Whether giant or not, these
components are characterized as follows (according to definitions
in \cite{Dorogovtsev01}):

\begin{itemize}
\item The weakly connected component, WCC, the percolative cluster
in undirected graphs. In the WCC, every vertex is reachable from
every other, provided that the directed nature of the edges is
ignored.

\item The strongly connected component, SCC, the set of vertices
reachable from its every vertex by a directed path.

\item The in-component, IN, all vertices from which the SCC is
reachable by a directed path.

\item The out-component, OUT, all vertices which are reachable
from the SCC by a directed path.
\end{itemize}

Notice that, with these definitions, the SCC in included in both
the IN and OUT components. The percolation theory developed by
Newman {\it et al.}\cite{Newman01,Callaway00} and Dorogovtsev {\it
et al.} \cite{Dorogovtsev01} for directed graphs with arbitrary
degree distribution and statistically uncorrelated vertices has
shown that there are two phase transitions: the one at which the
giant weakly connected component(GWCC) appears, and the one at
which the other three giant components appear simultaneously: the
giant in-component(GIN), the giant out-component(GOUT) and the
giant strongly connected component(GSCC) as the intersection of
the other two. The first phase transition corresponds in fact to
the standard phase transition in an undirected random graph
with arbitrary degree sequence and statistically uncorrelated
vertices. The condition for this phase transition was first given
by Molloy and Reed \cite{Molloy95} and reads
\begin{equation}
\sum_{k}k(k-2)P(k) \ge 0.
\label{eq:1}
\end{equation}
When this condition is fulfilled, and although some disconnected finite components
may remain, the GWCC emerges. It contains a macroscopic portion of
the vertices in the network capable to communicate to each other
regardless of the orientation of their links. The second phase
transition is characteristic of directed networks. The critical
point \cite{Newman01}
\begin{equation}
\sum_{k_{i},k_{o}}k_{o}(k_{i}-1)P(k_{i},k_{o}) = 0,
\label{eq:2}
\end{equation}
marks the first simultaneous appearance of the other three giant
components: the GSCC, the GIN and the GOUT, as well as other
secondary structures such as tubes or tendrils \cite{Broder00}.

\begin{figure}
\epsfig{file=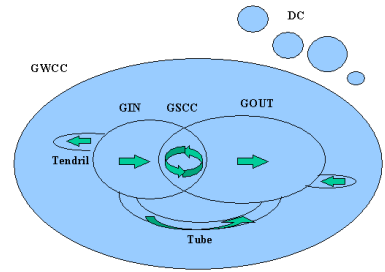,width=7cm}
 \caption{Schematic representation of the component structure of a directed network.} \label{figure1}
\end{figure}

The efforts to understand how this landscape is modified by the
consideration of correlations between degrees of neighboring
vertices has been exclusively focused on the percolation
analysis of undirected networks \cite{Newman022,Newman03,Vazquez03}. Assortative
mixing by degree, observed in the vast majority of social real
networks, has been found to favor percolation in the sense that
the giant component appears at lower edge density. On the
contrary, disassortative correlations, characteristic of
technological and biological networks, difficult the formation of
the giant component even if the second moment of the degree
distribution diverges.

\section{Generalized percolation}

In this section we develop  a general theory for percolation in
directed random networks with arbitrary two point degree
correlations and bidirectional edges. We term our networks
``random'' --or Markovian-- because, apart from purely local
properties and two point correlations being fixed, networks are
maximally random. This implies that the whole topology is encoded
into the degree distribution $P({\bf k})\equiv
P(k_{i},k_{o},k_{b})$ and the transition probabilities $P_{i}({\bf
k'} | {\bf k})$, $P_{o}({\bf k'} | {\bf k})$, and $P_{b}({\bf k'}
| {\bf k})$ measuring the probability to reach a vertex of degree
${\bf k'}$ leaving from a vertex of degree ${\bf k}$ using an
incoming, outgoing, or bidirectional edge respectively. Notice
that we have used the notation ${\bf k}\equiv(k_{i},k_{o},k_{b})$
and that we consider all three kind of edges as independent
entities. These transition probabilities are related through the
following degree detailed balance conditions \cite{Boguna02}
\begin{equation}
k_b P({\bf k}) P_{b}({\bf k'} | {\bf k})=k'_b P({\bf k'}) P_{b}({\bf k} | {\bf k'})
\end{equation}
and
\begin{equation}
k_o P({\bf k}) P_{o}({\bf k'} | {\bf k})=k'_i P({\bf k'}) P_{i}({\bf k} | {\bf k'}).
\label{detailed_dir}
\end{equation}
These conditions assure that any edge leaving a vertex points to
another vertex or, in other words, that the network is closed.
Notice that this may not be the case in situations where
information is incomplete and the out-degree of a vertex is known
but not the neighbors at the end of these edges.

To analyze the percolation properties for this class of networks,
it is necessary to calculate the joint distribution $G(s,s')$ of
the number of vertices (plus itself), $s$, that are reachable from
a given vertex and the number of vertices (plus itself), $s'$,
that can reach that vertex. Notice that we can leave a vertex using an outgoing or bidirectional edge and we can arrive to a vertex through an incoming or bidirectional edge. These sets of vertices are called the out- and in-components
of a given vertex, respectively. Analogously, we can define the
marginal probabilities $G_{o}(s)$ and $G_{i}(s)$ for the number of
reachable vertices from a given one and the number of vertices
that can reach it, respectively. These three functions contain the
information of the sizes of the different giant components of the
network. Notice that, if the network is above the percolation
threshold, $G_{o}(s) \ne \sum_{s'} G(s,s')$ and $G_{i}(s) \ne
\sum_{s'} G(s',s)$. The probability to belong to a finite
component is smaller than one in this situation, which implies
that the remaining corresponds to the probability to
belong to an infinite component, that is, the giant component.
Function $G_{o}(s)$, for instance, measures the probability that a given vertex has a finite out-component of size $s$ regardless of the size of its in-component, which can be finite or not. On the other hand, function $\sum_{s'} G(s,s')$ only accounts for finite components. Therefore, the relative sizes of the different giant components of the network can be written as
\begin{equation}
GOUT=1-\sum_s G_{i}(s) \mbox{ , } GIN=1-\sum_s G_{o}(s),
\label{gingout}
\end{equation}
and
\begin{equation}
GSCC=1-\sum_s G_{o}(s)-\sum_s G_{i}(s)+\sum_{s,s'} G(s,s'),
\label{gscc}
\end{equation}
where the GOUT is thought of as the set of vertices with an infinite
in-component, the GIN as the set of vertices with an infinite out-component, and the GSCC as the set of vertices with infinite in- and out-components simultaneously.

In heterogeneous networks, this set of probabilities depend on the
degree of the vertex from where we start the count. Therefore,
functions $G(s,s')$, $G_{o}(s)$, and $G_{i}(s)$ are to be expressed
as
\begin{equation}
G_i(s)=\sum_{\bf k} P({\bf k}) G_i(s|{\bf k}) \mbox{ , }  G_o(s)=\sum_{\bf k} P({\bf k}) G_o(s |{\bf k}),
\label{eq:7}
\end{equation}
and
\begin{equation}
G(s,s')=\sum_{\bf k} P({\bf k}) G(s,s'|{\bf k}),
\label{eq:8}
\end{equation}
where functions $G$ inside the summations have the same
meaning as the original ones under the condition of starting from
a vertex of degree ${\bf k}$. Thus, the complete solution of the
problem goes through finding the conditional
probabilities $G_i(s|{\bf k})$, $G_o(s|{\bf k})$ and $G(s,s'|{\bf
k})$. In the following subsections we will show how to compute
them in the general correlated case.

\subsection{In/Out component}

To proceed further, we first focus our attention on the
out-component size distribution of a vertex of degree ${\bf k}$,
$G_{o}(s|{\bf k})$. Starting from a vertex of degree ${\bf
k}=(k_i,k_o,k_b)$, we can leave it using the $k_o$ outgoing edges
and the $k_b$ bidirectional ones. Then, the number of reachable
vertices will be the sum of the reachable vertices of each of the
$k_o+k_b$ neighbors plus 1. In mathematical terms, this translates
into
\begin{widetext}
\begin{equation}
G_{o}(s|{\bf k})=\sum_{s_1}\cdots \sum_{s_{k_o+k_b}}
g_{o}(s_1|{\bf k}) \cdots g_{o}(s_{k_o}|{\bf k})
g_{o}^b(s_{k_o+1}|{\bf k}) \cdots g_{o}^b(s_{k_o+k_b}|{\bf
k})\delta_{s_1+\cdots+s_{k_o+k_b}+1,s}, \label{G_out(s|k)}
\end{equation}
\end{widetext}
where $g_{o}(s|{\bf k})$ ($g_{o}^b(s|{\bf k})$) is the
distribution of the number of reachable vertices from a vertex
given that we have arrived to it from another source vertex of
degree ${\bf k}$ following one of its outgoing (bidirectional)
edges. In writing Eq.(\ref{G_out(s|k)}), we have used the fact
that random networks are locally tree like.  Equations of the type
of (\ref{G_out(s|k)}) find in the discrete Laplace space their
natural representation in terms of the generating function
formalism. Using this formalism, these equations simplifies
enormously and can be manipulated very easily. Within this
formalism, equation (\ref{G_out(s|k)}) simplifies as
\begin{equation}
\hat{G}_{o}(z|{\bf k})=z \left[ \hat{g}_{o}(z|{\bf k})\right]^{k_o}\left[ \hat{g}_{o}^b(z|{\bf k}) \right]^{k_b},
\label{eq:10}
\end{equation}
where we have adopted the notation $\hat{f}(z)\equiv \sum_s f(s)
z^s$. In what follows, we will work in the discrete Laplace space,
using the generating function formalism. 

Functions $g_{o}$ and
$g_{o}^b$ satisfy the following set of coupled equations
\begin{equation}
\begin{array}{l}
\hat{g}_{o}(z|{\bf k})=z \displaystyle{\sum_{\bf k'}} P_{o}({\bf k'}|{\bf k})
\left[ \hat{g}_{o}(z|{\bf k'}) \right]^{k'_{o}} \left[ \hat{g}_{o}^b(z|{\bf k'}) \right]^{k'_{b}}\\[0.5cm]
\hat{g}_{o}^b(z|{\bf k})=z\displaystyle{\sum_{\bf k'}} P_{b}({\bf k'}|{\bf k})
\left[ \hat{g}_{o}(z|{\bf k'}) \right]^{k'_{o}} \left[ \hat{g}_{o}^b(z|{\bf k'}) \right]^{k'_{b}-1}
\end{array}
\label{transcendent_general}
\end{equation}

The term $k_b-1$ in the second line of Eqs. (\ref{transcendent_general}) comes from
the fact that one of the bidirectional edges has already been used
to reach the vertex of degree ${\bf k'}$ and, thus, cannot be used again
to leave it. Notice that this restriction is not needed in the first line of Eqs.
(\ref{transcendent_general}) since, in this case, we have reached
the vertex using an outgoing edge of the source vertex. The set of
equations (\ref{transcendent_general}) is closed for the functions
$g$. Its solution for $z=1$ will allow us to compute the size
of the GIN component,
\begin{equation}
GIN=1-\sum_{\bf k} P({\bf k})\left[
\hat{g}_{o}(1|{\bf k})\right]^{k_o}\left[ \hat{g}_{o}^b(1|{\bf k})
\right]^{k_b}.
\end{equation}

The trivial solution of Eqs. (\ref{transcendent_general}) is $\hat{g}_{o}(1|{\bf k})=1$, $\hat{g}_{o}^b(1|{\bf k})=1$, corresponding to the only case without giant component. Therefore, the network will percolate at the directed level when this trivial solution becomes unstable. To analyze the stability of this solution we use the
approach adopted in \cite{Vazquez03} and find solutions of the
form $\hat{g}_{o}(1|{\bf k})=1-\epsilon x({\bf k})$,
$\hat{g}_{o}^b(1|{\bf k})=1-\epsilon y({\bf k})$ in the limit
$\epsilon \rightarrow 0$. Replacing these expressions in Eqs.
(\ref{transcendent_general}) and taking the limit $\epsilon
\rightarrow 0$ we obtain
\begin{equation}
\left(
\begin{array}{c}
x({\bf k}) \\
y({\bf k})
\end{array}
\right)
=
\sum_{\bf k'} {\bf C}^o_{{\bf k}{\bf k'}}
\left(
\begin{array}{c}
x({\bf k'}) \\
y({\bf k'})
\end{array}
\right),
\end{equation}
where the matrix ${\bf C}_{{\bf k}{\bf k'}}^o$ is defined as
\begin{equation}
{\bf C}_{{\bf k}{\bf k'}}^o \equiv
\left(
\begin{array}{cc}
k'_oP_{o}({\bf k'}|{\bf k}) & k'_b P_{o}({\bf k'}|{\bf k})\\ [0.5cm]
k'_oP_{b}({\bf k'}|{\bf k}) & (k'_b-1)P_{b}({\bf k'}|{\bf k})
\end{array}
\right).
\end{equation}
The stability of the solution $\hat{g}_{o}(1|{\bf k})=1$,
$\hat{g}_{o}^b(1|{\bf k})=1$ is thus determined by the maximum
eigenvalue of the matrix ${\bf C}_{{\bf k}{\bf k'}}^o$,
$\Lambda_{m}$. When $\Lambda_{m} \le 1$ this solution is stable
and the GIN component does not exist. In contrast, when
$\Lambda_{m}>1$ a non trivial solution of the set of equations
(\ref{transcendent_general}) exists and the GIN component
emerges.

The analysis for the in-component of individual vertices is
identical to the case of the out one if we replace in equations (\ref{eq:10})
and (\ref{transcendent_general})
\begin{equation}
\begin{array}{rcl}
k_o & \rightarrow & k_i\\[0.3cm]
g_{o}(s|{\bf k}) & \rightarrow & g_{i}(s|{\bf k})\\[0.3cm]
g_{o}^b(s|{\bf k}) & \rightarrow & g_{i}^b(s|{\bf k}) \\[0.3cm]
P_{o}({\bf k'}|{\bf k}) & \rightarrow & P_{i}({\bf k'}|{\bf k})
\end{array}
\end{equation}
In this case, the matrix controlling the onset of the GOUT
component is
\begin{equation}
{\bf C}_{{\bf k}{\bf k'}}^i \equiv
\left(
\begin{array}{cc}
k'_iP_{i}({\bf k'}|{\bf k}) & k'_b P_{i}({\bf k'}|{\bf k})\\ [0.5cm]
k'_iP_{b}({\bf k'}|{\bf k}) & (k'_b-1)P_{b}({\bf k'}|{\bf k})
\end{array}
\right).
\end{equation}
As before, the condition for the appearance of the GOUT component
is ruled by the maximum eigenvalue of the matrix ${\bf
C}_{{\bf k}{\bf k'}}^i$. At first glance, one could be tempted to conclude
that, since matrices ${\bf C}_{{\bf k}{\bf k'}}^i$ and ${\bf
C}_{{\bf k}{\bf k'}}^o$ are different, their eigenvalues are also different,
leading to different phase transitions for the appearance of the GIN and GOUT
components. However, it can be proved that the eigenvalues
spectra of both matrices are identical and, then, both the GIN and the GOUT components appear
simultaneously.

It is illustrative to recover from this formalism the results for
the purely undirected and the purely directed cases. In indirected
networks, only bidirectional edges are present, $k_i \equiv 0$ and
$k_o \equiv 0$, and the matrices turns into
\begin{equation}
\begin{array}{rr}
{\bf C}_{{\bf k}{\bf k'}}^o \rightarrow & C_{k_b k'_b}=(k'_b-1)
P(k'_b|k_b)\\[0.3cm]
{\bf C}_{{\bf k}{\bf k'}}^i \rightarrow & C_{k_b k'_b}=(k'_b-1)
P(k'_b|k_b),
\end{array}
\end{equation}
recovering results in \cite{Vazquez03}. In the case of purely directed
networks, $k_b \equiv 0$ and we obtain
\begin{equation}
\begin{array}{rr}
{\bf C}_{{\bf k}{\bf k'}}^o \rightarrow & C_{{\bf k}{\bf
k'}}^o=k'_o
P_o({\bf k'}|{\bf k})\\[0.3cm]
{\bf C}_{{\bf k}{\bf k'}}^i \rightarrow & C_{{\bf k}{\bf
k'}}^i=k'_i P_i({\bf k'}|{\bf k}).
\end{array}
\end{equation}
This result generalizes the percolation theory for purely directed random networks developed in  \cite{Newman01,Callaway00,Dorogovtsev01} to the case of arbitrary degree-degree correlations.

\subsection{Strongly connected component}

The analysis of the GSCC requires a more careful development of the ideas
introduced in the previous section. In this case, the joint distribution of
the in- and out-components of a vertex of degree ${\bf k}$ reads
\begin{equation}
\hat{G}(z,z'|{\bf k})= z z'   \left[ \hat{g}_{o}(z|{\bf k})\right]^{k_o}\left[ \hat{g}^b(z,z'|{\bf k}) \right]^{k_b}  \left[ \hat{g}_{i}(z'|{\bf k})\right]^{k_i},
\end{equation}
where the function $g^b(s,s'|{\bf k})$ is defined analogously
 to $g_{o}^b(s|{\bf k})$ and $g_{i}^b(s|{\bf k})$. It is worth to
mention that, if the network contains bidirectional edges,
$G(s,s'|{\bf k})\neq G_o(s|{\bf k})G_i(s'|{\bf k})$ because, in
this case, such edges are common to the in and out components of
the vertex. Using the same reasoning as in the previous section,
we can write down a closed equation for the joint distribution
$g^b(s,s'|{\bf k})$
\begin{widetext}
\begin{equation}
\hat{g}^b(z,z'|{\bf k})= z z'  \sum_{{\bf k'}} P_b({\bf k'}|{\bf k}) \left[ \hat{g}_{o}(z|{\bf k'})\right]^{k'_o}\left[ \hat{g}^b(z,z'|{\bf k'}) \right]^{k'_b-1}  \left[ \hat{g}_{i}(z'|{\bf k'})\right]^{k'_i}.
\end{equation}
\end{widetext}
This equation, together with Eq. (\ref{eq:10}) and Eqs. (\ref{transcendent_general}) are the complete solution of the problem.
The solutions for arbitrary values of $z$ and $z'$ allow to find the distribution
of the sizes of the in- and out-components of single vertices whereas the
non-trivial solution for $z=1$ and $z'=1$, combined with Eqs. (\ref{gingout}-\ref{gscc}) and Eqs. (\ref{eq:7}-\ref{eq:8}), will provide us with a method to calculate the size of the different giant components of the network.

\section{Uncorrelated networks}

As mentioned before, real networks are usually correlated in the
sense that the degrees of pairs of connected vertices are
correlated random quantities. Nevertheless, uncorrelated networks
are equally useful as benchmarks, or null models, to test
topological and dynamical properties and compare them to the
results obtained in correlated networks. 

When two point
correlations are absent, the transition probabilities become independent
of the degree of the source vertex. In this situation, after some
elemental algebra, we obtain
\begin{equation}
 P_{o}({\bf k'} | {\bf k})=\frac{k'_i P({\bf k'})}{\langle k_i \rangle} \mbox{ , }
 P_{i}({\bf k'} | {\bf k})=\frac{k'_o P({\bf k'})}{\langle k_i \rangle}
\label{transitioninout_uncorrelated}
\end{equation}
and
\begin{equation}
P_{b}({\bf k'} | {\bf k})=\frac{k'_b P({\bf k'})}{\langle k_b
\rangle}, \label{transitionbi_uncorrelated}
\end{equation}
where we have made use of the fact that $\langle k_o \rangle =
\langle k_i \rangle$. Using these expressions, the set of
equations (\ref{transcendent_general}) reduces to the following
set of trascendent equations,
\begin{equation}
\begin{array}{ll}
y=\langle k_i \rangle^{-1} \partial_x \hat{P}(1,y,z) &
z=\langle k_b \rangle^{-1} \partial_z \hat{P}(1,y,z)\\[0.5cm]
x=\langle k_i \rangle^{-1} \partial_y \hat{P}(x,1,z')&
z'=\langle k_b \rangle^{-1} \partial_{z'} \hat{P}(x,1,z')\\[0.5cm]
z''=\langle k_b \rangle^{-1}\partial_{z''} \hat{P}(x,y,z''),
\label{transcendent}
\end{array}
\end{equation}
where $\hat{P}(x,y,z)$ is the generalized generating function of the degree distribution, that is,
\begin{equation}
\hat{P}(x,y,z)\equiv \sum_{k_i,k_0,k_b} x^{k_i} y^{k_o}z^{k_b} P(k_i,k_o,k_b).
\end{equation}
The condition for the existence of a non trivial solution of the
set of equations (\ref{transcendent}) is easily obtained using the
formalism developed in the previous sections. For the uncorrelated
case, the matrices ${\bf C}_{{\bf k}{\bf k'}}^i$ and ${\bf
C}_{{\bf k}{\bf k'}}^o$ become independent of ${\bf k}$ and its
maximum eigenvalue reads
\begin{widetext}
\begin{equation}
\Lambda_m=\frac{1}{2}\left\{\frac{\langle k_b(k_b-1)\rangle}{\langle k_b \rangle}+\frac{\langle k_i k_o \rangle}{\langle k_i \rangle}+\sqrt{\left(\frac{\langle k_b(k_b-1)\rangle}{\langle k_b \rangle}-\frac{\langle k_i k_o \rangle}{\langle k_i \rangle}\right)^2+\frac{4 \langle k_i k_b \rangle \langle k_o k_b \rangle}{\langle k_i \rangle \langle k_b \rangle}} \right\},
\label{eigenvalue_uncorrelated}
\end{equation}
\end{widetext}
so that, whenever the condition $\Lambda_m \ge 1$ is fulfilled, the network is in the percolated phase.
As it can be seen from Eq. (\ref{eigenvalue_uncorrelated}), the presence of bidirectional edges alters
the point at which the giant component arises in a non trivial way.
The term $\Gamma=\langle k_i k_b \rangle \langle k_o k_b \rangle/\langle k_i \rangle \langle k_b \rangle$
measures the strength of the coupling between directed
and bidirectional edges. When this coupling is weak, $\Gamma \sim 0$ and the maximum eigenvalue takes the simple form
\begin{equation}
\Lambda_m\sim \max \left\{ \frac{\langle k_b(k_b-1)\rangle}{\langle k_b \rangle}, \frac{\langle k_i k_o \rangle}{\langle k_i \rangle} \right\}.
\end{equation}
This result is easy to understand since, when $\Gamma \sim 0$, vertices cannot have directed and
bidirectional edges simultaneously and, as a consequence, the network is composed of two isolated
networks, one of them purely directed and the other one containing bidirectional edges only.

In the purely undirected case, $k_i \equiv 0$ and $k_o \equiv 0$
and we recover the well known condition for percolation in
undirected networks with given degree distribution Eq. (\ref{eq:1})
\begin{equation}
\Lambda_m=\frac{ \langle k_b(k_b-1) \rangle}{\langle k_b \rangle}
> 1.
\end{equation}
In the case of purely directed networks, $k_b \equiv 0$ and the
maximum eigenvalue reads
\begin{equation}
\Lambda_{m}=\frac{\langle k_i k_o \rangle}{\langle k_i \rangle} >
1,
\end{equation}
recovering Eq. (\ref{eq:2}).

When $\Gamma \gg 1$, $\Lambda_m$ must be computed using
Eq.(\ref{eigenvalue_uncorrelated}) and, in general, will depend on
the density of edges, as well as on the type of correlations
between directed and bidirectional edges. In particular, a
positive correlation between $k_b$ and $k_i$ or $k_o$ can strongly
favor the emergence of the giant component even if the density of
bidirectional edges is very small. We will illustrate this point
in the example of the next section.

To finish the uncorrelated analysis, let us compute the relative
sizes of the giant components. Let $(x_c,y_c,z_c,z'_c,z''_c)$ be the non
trivial solution of the set of Eqs. (\ref{transcendent}), then using Eq. (\ref{gingout}) and Eq. (\ref{gscc}),
the relative sizes of the different giant components of the network read
\begin{equation}
GIN=1-\hat{P}(1,y_c,z_c) \mbox{  ,  } GOUT=1-\hat{P}(x_c,1,z'_c)
\end{equation}
\begin{equation}
GSCC=1-\hat{P}(1,y_c,z_c) -\hat{P}(x_c,1,z'_c)+\hat{P}(x_c,y_c,z''_c)
\end{equation}

In the next section, we present a practical application of this
formalism.

\section{Bidirectional edges as percolation catalysts}

\begin{figure}
\epsfig{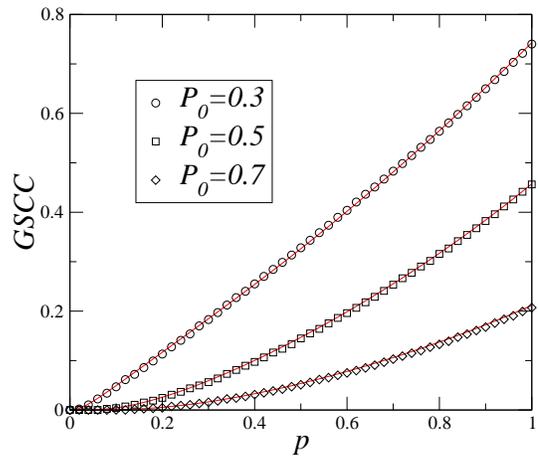}
 \caption{Relative size of the giant strongly connected component as a function of the conversion probability $p$. Simulation results are for a single network with $N=10^6$ vertices. Solid lines are the numerical solution of the set of equations (\ref{transcendent}).} \label{GSCC}
\end{figure}

Suppose we have a purely directed network with degree distribution $P(k_i,k_o)$
and no two point degree correlations. Suppose also that the network is in a regime
in which the GWCC exists but not the GSCC, that is,
\begin{equation}
\frac{\langle k_i k_o \rangle}{\langle k_i \rangle}<1 \mbox{    and    } \frac{\langle (k_i+k_o)(k_i+k_o-1)\rangle}{\langle (k_i+k_o)\rangle} > 1.
\end{equation}
Now we transform the original network by converting each directed edge into
a bidirectional one with probability $p$. After this transformation, we end up with
a network with $p E$ bidirectional edges and $(1-p)E$ directed ones, where $E$ is the
original number of directed edges on the network. The degree distribution of the
transformed network can be written, in the discrete Laplace space, as
\begin{equation}
\hat{P}_p(z_i,z_o,z_b)=\hat{P}(pz_b+(1-p)z_i,pz+(1-p)z_o).
\label{pconjuntafinal}
\end{equation}
This transformation undergoes a phase transition as we increase the value of $p$.
When $p=0$, the network is purely directed and, by construction, it has no GSCC.
When $p=1$, all edges become bidirectional and, thus, the GSCC is identical to the GWCC
of the original network. Therefore, at some intermediate value $p_c$, the network percolates
and a GSCC emerges. The value of $p_c$ can be easily obtained using the expression for
the maximum eigenvalue, Eq. (\ref{eigenvalue_uncorrelated}), and the final degree distribution, Eq.(\ref{pconjuntafinal}).

\begin{figure}
\epsfig{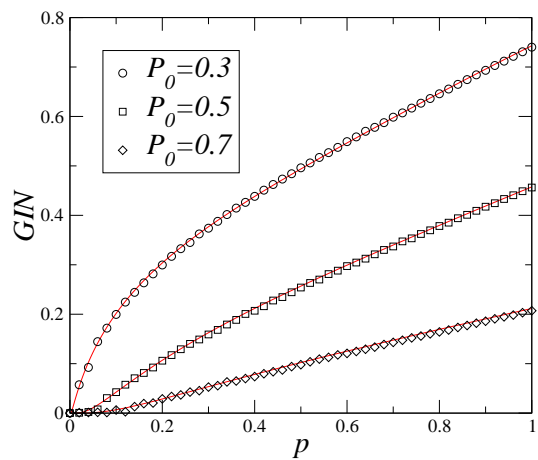}
 \caption{Relative size of the giant in component as a function of the conversion probability $p$. Simulation results are for a single network with $N=10^6$ vertices. Solid lines are the numerical solution of the set of equations (\ref{transcendent}).} \label{GIN}
\end{figure}

The most interesting case corresponds to networks with marginal degree distributions
following power laws of the form $P_i(k_i)\sim k_i^{-\gamma_i}$ and $P_o(k_o)\sim k_o^{-\gamma_o}$
with $\gamma_i,\gamma_o \le 3$. When the transformation described above is performed in this type
of networks, some of the terms in Eq.(\ref{eigenvalue_uncorrelated}) are
proportional to $\langle k_i^2 \rangle$ and $\langle k_o^2 \rangle$ and,
consequently, $\Lambda_m \rightarrow \infty$ in the thermodynamic limit. This, in turn, implies
that $p_c=0$, that is, even an infinitely small fraction of bidirectional edges suffices to percolate the network.

\subsection{Numerical simulation}

The check the accuracy of our theory in the case of power law
marginal degree distributions, we have performed extensive
numerical simulations. We first generate purely directed random
networks with degree distribution of the form
$P(k_i,k_o)=P_i(k_i)P_o(k_o)$ and no two point correlations. The
in and out degree distributions are taken to be identical and to
follow a scale-free form of the type
\begin{equation}
P_i(k)=P_o(k)=
\left\{
\begin{array}{lr}
P_0 & k=0\\[0.5cm]
\displaystyle{\frac{(1-P_0)}{\zeta(\gamma) k^{\gamma}}} & k \ge 1
\end{array}
\right.
\label{eq:33}
\end{equation}
where $\zeta(\gamma)$ is the Zeta Riemann function. To generate a purely directed random network, we use a natural
extension of the configuration model
\cite{bekessi72,benderoriginal,Molloy95,Molloy98} --an
algorithm intended to generate uncorrelated random networks with a
given degree distribution. The algorithm starts by first
assigning to a set of $N$ vertices a pair of ``stubs'', one of
them incoming and the other one outgoing, $k_i$ and $k_o$,
randomly drawn from the distribution $P(k_i,k_o)$. The only
requirement is that $\sum_{i}k_i=\sum_i k_o$, whenever one wishes
to close the network. The network is constructed by selecting
pairs of in and out stubs chosen uniformly  at random to create
directed edges, avoiding multiple, bidirected, and
self-connections among vertices. Once the network has been
assembled, each directed edge is transformed into a bidirectional
one with probability $p$, and the relative sizes of the giant
components are measured.

Figures \ref{GSCC} and \ref{GIN} show simulation results for a
scale-free network following Eq.(\ref{eq:33}) with exponent
$\gamma=3$ and size $N=10^6$ as compared to the numerical solution
of Eqs.(\ref{transcendent}). As it can be seen, the agreement
between simulation results and the theoretical prediction is
excellent. The relative sizes of the GSCC and the GIN are shown as
a function of the conversion probability $p$ for different values
of the distribution parameter $P_0$, the probability of nodes
having null in or out degree. Even for very small values of $p$,
the GSCC and the GIN are evident. As expected, small values of
$P_0$ favor the growth of bigger giant components. Then,
bidirectional edges act as a percolation catalysts, favoring the
appearance of a fine structure in the giant connected component. The
scale-free property of many real networks is, once more, indicative
of interesting features, since, in this case, the presence
of an infinitesimal fraction of bidirectional edges is enough to
ensure percolation at the level of the directed components .

\section{Conclusions}

We have derived a very general formulation of the theory of percolation in directed random networks with bidirectional edges and arbitrary two point degree correlations. Our formalism accounts for all the previously known results for percolation in purely directed and purely undirected random networks, which stand as limiting cases of our theory. The percolation threshold for the most general situation is derived as a function of the maximum eigenvalue of the connectivity matrices. In particular, for networks with no two point correlations, explicit expressions are provided depending on the first and second moments of the degree distribution $P(\bf{k})$. In this case, we have also shown that bidirectional edges act as a catalyst  for percolation, favoring the emergence of the GSCC, and for scale-free networks, only an infinitesimal fraction of bidirectional edges is needed.

After the completion of this work, we have become aware of a recent preprint \cite{Meyers04} where the classical Susceptible-Infected-Recovered (SIR) model of epidemiology is analyzed in uncorrelated directed random networks with bidirectional edges. Since there exist a mapping between the SIR model and percolation theory, some of the results derived in that reference overlap our results for the uncorrelated case.

\begin{acknowledgments}
We would like to thank R. Pastor-Satorras for valuable comments. This work
has been partially supported by DGES of the Spanish government,
Grant No. FIS2004-05923-CO2-02, and EC-FET Open project COSIN
IST-2001-33555. M. B.  acknowledges financial support from the
MCyT (Spain).
\end{acknowledgments}

\end{document}